# EFFECT OF SOCIABILITY AND CURIOSITY OF SENIOR DEVELOPERS IN BUILDING AGILE SCRUM TEAM COMPETENCY

Ravi Kalluri

Department of Engineering Management & Systems Engineering,
Old Dominion University, Norfolk, Virginia, USA

*ABSTRACT*

*This paper aims to investigate the mechanisms that contribute to propagation of competence in an Agile Scrum team. This study seeks to challenge the traditional view of bounded rationality (BR). An Agile Scrum team (Team) is expected to build problem solving competence quickly as the expected ramp up time continues to shrink. But the team has a mixture of expertise, competence and sociability levels that affect out-of-the-box performance. The objective is to expand BR into the social realm and see how teams can self-organize and reconfigure to allow effective problem solving. Studies have shown that agent-based computational simulation is an appropriate technique to explore this point from a theoretical perspective. (Fioretti, 2013) (Secchi, 2015). The first step is to define the problem, discuss how senior team members exhibit high curiosity and apply sociability and cognitive resources to develop overall team competence. This dynamic is modeled and simulated in NetLogoR and the results are analyzed. Finally, some key findings are presented and discussed.*

## 1. INTRODUCTION

This study explores how the curiosity and sociability traits, if present in the senior developers of an Agile Scrum team, can improve team wide problem-solving competence. This paper introduces a refined version of Bounded Rationality (BR) that is socially oriented (Simon, 1993). In particular, it makes the distinction between senior developers who operate within the boundaries of their team (i.e. sociable) and those who extend their reach outside of those boundaries and norms (i.e. curious). An existing agent-based simulation model is repurposed where agents represent developers and user stories. A key trait for agent developer is competence while a key trait for agent story is difficulty.

The curiosity and sociability traits of a senior developer are expected to combine team member competences in a way that makes problem solving more efficient. This is due to the understanding that the integration of knowledge from the various team members is done non-linearly, i.e. exponentially increasing or decreasing the original knowledge base of the curious senior developers. Conversely, other non-curious senior developers would combine competences in a linear fashion, and this makes knowledge integration more directly dependent on the existing competences (preferred to be at higher level). In other words, curiosity may add an emergent element to team competence that is not included in the original knowledge base of each team member. This synergy has the potential to dramatically increase the efficiency of knowledge integration and resilience in an Agile Scrum team when dealing with complex problems.

The paper shows how the traits of sociability and curiosity enable senior members in an Agile Scrum team to collaborate with junior team members in completing user stories and in the process raise the overall team competence. It shows that BR can be considered as a social process and the simulation of team dynamics reveals how social attitudes can aid in problem solving.

The paper is structured as follows. Section 1 provides an introduction. Section 2 presents the theoretical background and literature review relevant to the study, supporting reasons for inquiry, and contextual background. Section 3 discusses the Agent Based Simulation model. Section 4 describes the procedures, data analysis and simulation results. Section 5 discusses the implications, future work and concludes the paper. This is followed by references and an appendix.

## 2. RELATED WORK

### 2.1. Bounded Rationality (BR)

Simon (1955, 1997) posited that rationality is always bounded in the sense that a decision is made regardless of all the alternatives that are available. That is because of the inherently limited alternatives available in any given situation (March, 1994) (Simon, 1997). The limitations forced upon our rationality by our inherent boundedness prevent us from selecting the optimal set of means (Secchi, 2016). This limits our ability to effectively solve complex problems. The result is we end up creating new problems. For example, figuring out how to increase cooperation in a team is a complex goal. According to Simon, it is not possible to take the goal and work backwards to derive the sequence of the right means, as the traditional view on BR would assume. The manner in which the team members need to engage with each other and with the environment has to be atypical in order to find a potential solution. (Chia and Holt, 2009) (Secchi, 2016).

Simon (1955, 1997) suggests that engagement with the environment is central to BR as a source of resources. There is a tendency exhibited by certain agents (members of a team) who rely more on gaining competency through social channels by watching the senior team members. The context of "social channels" has changed with the advent of the internet (Secchi, 2011) (Magnani et al., 2007). Competency can be gained by direct consultation with a team member or by technological instruments (i.e. smartphones) or services (i.e. social media like Facebook or Twitter, Google search, etc.).

A senior developer (senior team member) who is sociable, prefers to make decisions assuming the existence of a social group to refer to. (Bardone, 2011) (Secchi and Bardone, 2009, 2013). Therefore, sociable individuals intrinsically take a more collaborative stance on decision-making. When this sociability trait is exhibited by the senior team members, there is rapid diffusion of competence to the novice team members. This enables faster build-up of overall team effectiveness in solving problems (completing user stories).

Lalsing (2012) studied three different sized Agile teams developing products based on the same technologies and using Scrum. Both objective and subjective measures were used and the results are supported by a survey. The results clearly show that for agile methodologies to work well, it is crucial to select the "right" people for the right team.

However, the variety of ways in which an agent (sociable, senior developer) may actually engage with the other team members is not fully understood. The impact of sociability is well substantiated when interactions within a group are stable and well-defined, but it is unclear how much this concept is useful when groups are formed ad hoc or when the decision maker reaches out to members of other groups (the environment).

The motivation for creation is higher for the agent as it leads to competency gain and external recognition. The internal sharing of information, on the other hand, develops the team as a whole but eats into the time for personal advancement and also narrows the competency gap with more junior developers. With such complex psychological forces at work, it is hard to assess whether the notion of sociability can really have an impact on the way in which bounded agents may really act. There is thus a need to define a more inclusive term "curiosity" for the specific type of engagement that the agent has with the environment and the team. A curious agent is continuously learning by enquiry and openly exploring ideas and decisions with the team and environment using the available social channels.

The notion of curiosity is not aligned with the extant view on BR because it prompts the agent to learn from open-ended explorations of complex problems and to accept the complexity of decision-making. Curiosity has been referred to as an "openness to cognitive diversity" (Klein and Kozlowski, 2000) and "cross-understanding" (Meslec and Graff, 2015). Curiosity allows the team member (agent) to cross existing boundaries of cooperation to form new knowledge associations for both personal and team interests.

## 2.2. Sociability and Curiosity

Sociability places importance on the information provided by social channels and an inclination for individuals to share ideas with like-minded people and collaborate. (Simon, 1993) (Knudsen, 2003). But sociability refers to situations in which a person works with other people on something that is mostly defined. Thomsen (2016) explains this concept with teams of medical doctors and nurses in the emergency room of a hospital. The ER team members are sociable individuals as they tend to work within boundaries that are set beforehand. This is a worldview specifying already accepted templates of thinking along with the identification of specific and well-defined problems and issues to deal with. Sociability can lead to formation of tight couplings among team members and, possibly, an entire organization but there is a risk that the organization becomes unfit to learn and adapt to the changing external environment (Rivkin, 2000), reaching what Siggelkow and Rivkin (2005) call "sticking points".

Curiosity, on the other hand, breaks preset patterns of behavior and allows individuals to see unexpected connections among apparently unrelated things (Bardone, 2011). From a social point of view, curious individuals reach out to others to explore problems more broadly (beyond the immediate team) and facilitate a solution (Secchi, 2011). The way these individuals interact with others is oriented toward gaining a better understanding of the problem at hand. Their use of information is not simply the sum of what is available from others but a restructuring of available expertise to find the best solution. The focus is on gaining new knowledge and understanding by interacting with the social environment. If sociability allows self-organization to emerge within a team, curiosity has the potential to establish new consortiums, both within and outside the team. This is necessary to build a resilient team. This is in contrast with the literature on "shared cognition" (Cannon-Bowers et al., 1993) (Cannon-Bowers and Salas, 2001) (McAvoy and Butler, 2007), where individuals share a mental model or group think to make the team more effective. Curiosity requires openness to learning and engaging with other people regardless of their background, position and role within an organization and may lead to better problem-solving. From this perspective, curiosity serves as a catalyst to sociability and makes team self-organization and resilience possible. The next section discusses how the sociability enhanced by curiosity of senior team members (agents with high competence) helps in building overall team competency and eventually productivity, self-organization and resilience in an Agile Scrum team with the aid of an agent-based simulation (ABS) model.

## 3. THE ABS MODEL

Curiosity is seen as a tool that links agents together; hence, it affects how teams deal with problems. Agent-based modeling (ABM) is a simulation technique that has been increasingly used in the social sciences (Secchi and Neumann, 2016) (Fioretti, 2013) and its properties have recently been explored in relation to teams (Secchi, 2015).

### 3.1. Objectives of the Model

The purpose of this model is to understand the impact of sociability and curiosity on boundedly rational agents that are confronted with complex problems. As mentioned above, curious senior developers reach out to others within and outside the team to explore problems more broadly and facilitate a solution. The way these individuals interact with others is oriented toward gaining a better understanding of the problem at hand. Their use of information (competence, c in the model) is not simply the sum of what is available from others but a deliberate reassembly of knowledge with the aim of finding the most effective solution. Given these assumptions, the model attempts to explore whether individuals with sociability and curiosity deal with problems better than individuals with sociability but lower curiosity, thus offering a better way to build overall team competence. The model mimics an organizational environment, where simulated developers (agents) have to deal with problems, and Agile teams are swarming around problems when one simulated developer cannot solve the problem alone. This represents an opportunity to explore how ad hoc teams are formed and how the sociability and curiosity behaviors of the senior team members influence competency building in the team.

### 3.2. Agents and Parameters

The model has two separate types of agents: stories (st) and developers (sd). The stories are characterized by level of difficulty (d) which is a random-normal value. The developers have a level of competence c that is also a random-normal value. The competence level c is an agent's knowledge, and it can be applied to a problem as a direct function of the efforts necessary to find a solution. In other words, c can be thought of as a set of cognitive abilities that each and every individual has in line with the tradition of studies on BR (Kahneman, 2003) (Gigerenzer and Selten, 2001). These are operational abilities that aid in problem solving. If $c > d$, then a problem gets solved, and competence of the agents responsible for finding a good solution increases at a fixed rate ri in [0.15, 0.30]. When a solution cannot be found, then the competence rate decreases at a fixed rate rd in [0, 0.05, 0.1]. This information is updated at every step of the simulation. All parameters and respective values are summarized in Table 1. A difficulty level d for each problem is not taken as an objective value because it depends on competence level c, the combination of agents around a given problem and the inclination of the agents to share and combine c based on their sociability and curiosity levels.

Table 1. Parameter notations and values.

| Parameter | Values | Description |
|---|---|---|
| Steps | 10 | The maximum number of opportunities that developers have to interact with each other when working on stories |
| Initial number of user stories, $N_{st,0}$ | 50, 100, 200 | Initial number of user stories in the product backlog at time zero |
| Problem spin-off, $pso$ | 2, 4, 10 | This is the factor by which stories can increase at any step of the simulation |
| Initial number of developers, $N_{sd,0}$ | 50, 100, 200 | Initial number of developers in an Agile Scrum team at time zero |
| Difficulty, $d$ | $\sim N(3, 1)$, $\sim N(1, 1)$ | Each story is associated with a difficulty level that is random-normally distributed |
| Competence, $c$ | $\sim N(1, 1.5)$, $\sim N(2, 1.5)$, $\sim N(3, 1.5)$ | This is the expertise of each developer that is needed to complete a story |
| Competence increase rate, $r_i$ | 0.15, 0.30 | The rate at which competence increases when a story is completed |
| Competence decrease rate, $r_d$ | 0, 0.05, 0.1 | The rate at which competence decreases when a story is not completed |
| Sociable developer, $sosd$ | $\sim N(0, 1)$ | This is the sociability of each developer and inclination on information coming from others in the team |
| Sociable | true, false | This influences the ways of working of the developers in solving problems in a team |
| Enquiry, $e$ | $\sim N(0, 1)$ | This is the enquiry level that would make developers consider knowledge coming from others in the environment |
| Curious | true, false | This influences the ways of working of the developers in solving problems in a team |
| Proximity | [0, 20, 1] | This is the value used to explore the environment that surrounds each agent |

The other characteristics of developers (sd) are enquiry e and socially oriented decision making sosd or sociability. Developers (sd) with sosd < Mean(sosd) - 0.75·Stdev(sosd) are less prone to use information from social channels to make decisions. On the other hand, those with sosd > Mean(sosd) + 0.75·Stdev(sosd) are particularly inclined on using information from social channels (other agents in the system) to make decisions. sosd models the behavior of developers in a team with different dispositions toward giving and taking information, recommendations, advice from others. Enquiry e is assigned to each agent using a random normal distribution. When the level ei of a particular agent i is higher than the mean e, then there are higher gains from cooperating with others. While high sosd indicates the ability to learn from others, e makes the developer explore solutions outside of the team. sosd*c is the amount of competence that a developer is willing to share with others without sacrificing individual growth and recognition.

Each developer (sd) scans the environment around it upto a pre-specified range and takes on stories along with other agents (team members) within that range. The developers then start sharing their knowledge (c), using different rules based on their sosd and e levels. Higher values of sosd imply better access to one's competence c. Also, developers with higher levels of sosd embed other's knowledge according to a non-linear effect, given by $\sum jci + (sodmj \cdot cj)sodmi \cdot ci$, where the parameters indexed with i refer to the agent and those indexed with j refer to other agents in range and connected to the agent i. Those with lower levels of sodm use a linear effect, $\sum jci + (sodmj \cdot cj)$. The assumption of this model is that curiosity triggers developers to process information coming from social channels in a way that combines their competence with the competence of others producing effects that may be valuable to problem solving. In the simulation, it is possible to switch curiosity "on" and "off" to understand how it impacts group decision-making and whether it enables more stories to be completed. The non-linear effect operates in the model for high e and sosd agents only when curiosity is turned "on".

Stories and developers with varying level of competence appear randomly in the simulated environment. While the stories do not move at all in the simulation space, the developers move around and try to find stories to complete. The developer's task is to find stories and swarm a team of other developers in the proximity around that story if he/she cannot solve it solely with own competence level c. When connection between a story and a team is established, the team of developers does not move until that story is done. Open stories are shown in green color. Done stories are shown in blue color. Developers with high competence are depicted in red color, intermediate developers are in yellow color and junior developers are in white color.

### 3.3. Process Overview

There are three steps in the simulation process. First, the agents (stories and developers) are randomly distributed on the two-dimensional simulation space (patch) and are attributed the characteristics described above. Second, the developers find stories, connect to them and to other developers. Third, senior developers try to complete the story with their own knowledge $ci$, and, if that is not enough, they combine knowledge from others in the team $cj$ according to curiosity levels (or its absence). We refer to those rules as linear and non-linear effects, the latter being those for more curious agents. Also, sd agents increase or decrease their competence and eventually change their general attitudes toward problem solving (sosd), depending on the results of the previous round of interactions. This can be called behavior emergence of senior developers in the team and is critical to competency building in a team.

## 4. PROCEDURES AND RESULTS

### 4.1. Procedures

The model was implemented in NetLogo 6.2.2, a free software for ABM (Wilensky, 1999). Guidance was taken on how many times each simulation scenario should run (Secchi and Gullekson, 2016) (Radax and Rengs, 2010). An attempt to reach a power of 0.95 at the 0.01 significance level and a conservative estimate of effect size, 0.1, resulted in approximately 30 runs for each simulation. The simulations were run in BehaviorSpace, a software tool integrated with NetLogo that allows us to perform experiments with models. BehaviorSpace runs a model many times, systematically varying the model's settings and recording the results of each model run. This process is sometimes called "parameter sweeping". It lets us explore the model's "space" of possible behaviors and determine which combinations of settings cause the behaviors of interest. In my computer with eight processor cores, eight simulations could be run in parallel.

### 4.2. Results

The main objective of the ABM simulation is to understand the effect of sociability and curiosity of senior developers (those with high problem-solving competence) on building the overall competency of an Agile Scrum team. For this purpose, all other variables in the model were held constant throughout the experiment as shown in Table 2.

Table 2. List of Constants.

| Variable | Value |
| --- | --- |
| Initial # stories (st) | 100 |
| Initial # developers (sd) | 50 |
| mean_enquiry | 0 |
| increase_comp_rate | 0.13 |
| tolerance | 5.5 |
| stdev_enquiry | 5.1 |
| decrease_comp_rate | 0.64 |
| mean_difficulty | 5 |
| stdev_difficulty | 5.2 |
| avoid-edges | TRUE |
| stdev_soc-or-sd | 5.03 |
| mean_soc-or-sd | 0.2 |
| looking_for_stories | TRUE |
| proximity | 19 |
| mean_competence | 5.0 |
| stdev_competence | 9.7 |

The simulations involved varying the two variables sociable and curious in combination leading to the following four scenarios:

- Scenario 1: (Sociable = F; Curious = F)
- Scenario 2: (Sociable = T; Curious = F)
- Scenario 3: (Sociable = F; Curious = T)
- Scenario 4: (Sociable = T; Curious = T)

Results of the NetLogo 6.2.2 simulation of the four scenarios are presented in Table 3.

Table 3. Simulation Results.

| Variable | Scenario 1 (Sociable = F; Curious = F) | Scenario 2 (Sociable = T; Curious = F) | Scenario 3 (Sociable = F; Curious = T) | Scenario 4 (Sociable = T; Curious = T) |
|---|---|---|---|---|
| Initial # stories (st) | 100 | 100 | 100 | 100 |
| Initial # sr. devs (sd) | 50 | 50 | 50 | 50 |
| # stories completed (st.solved2) (median) | 98 | 98 | 97 | 97 |
| Time steps (t) (median) | 4 | 3 | 4 | 2 |
| Team velocity (#stories completed / time) (median) | **28** | **25** | **32** | **46** |
| Max # of stories completed by | Mid-level dev | Mid-level dev | Senior dev | Junior dev |
| Highest competence gained by (median) | Junior dev | Senior dev | Junior dev | Junior dev |

Number of runs for each scenario (n) = 30

All scenarios use the team velocity (number of stories completed per unit time) as the dependent variable and differ from each other in the sociable and curious binary values. The values listed are the median values from 30 simulation runs for each of the four scenarios leading to a total of 120 runs. This was done to sufficiently account for the variability in the run results. The median values can thus be considered as reliable indicators of the behaviors under study.

The team velocity is generally positively correlated with the condition of curiosity, that makes senior developers with high competence (red sd's) enhance their capacity to interact with and develop the less competent team members (yellow and white sd's). In the simulation, curious is a Boolean variable, and it can be "true" (on) or "false" (off). The anchor for the results in the model is the "false" condition, meaning that we observe what happens to the dependent variable when curiosity is "true". The strongest effect to team velocity (46) is observed when both sociable and curious are set to "true". When sociable is set to "false" and curious is still "true", the team velocity significantly lower (32). It is apparent that sociability enhances the effect of curiosity and has a marginal although stable direct effect on team velocity.

Curiosity is the parameter that makes senior developers non-linearly combine efforts of the team members. This implies that the way to combine competences together in a team is more efficient when curiosity exists. This is evidenced by the fact that the curve for the number of (open) stories declines more rapidly when curiosity is "on".

Another interesting effect reported is the gain in competence of team members by initial competence level. In all but one scenario, junior developers show the largest growth in competence as a result of learning from their senior peers and observing them complete stories. But it needs both sociable and curiosity for the junior developers to translate this enhanced competence into a commensurate increase in velocity. Also, sociability of senior developers is alone not sufficient to percolate their competence beyond mid-level developers. In the absence of curiosity and a lack of tendency to learn from each other and other Agile Scrum teams in the organization environment, sociability without curiosity does not grow any significant team wide competence. The result also narrates something about the attachment preference of senior developers – the preference is to attach with other senior developers followed by mid-level developers and only in the end with junior developers. This might be revealing a natural inclination of senior developers towards individual vs. team performance because attaching to high competency developers may lead to more completed stories and hence more senior developer velocity. This natural tendency is mitigated when senior developers exhibit both sociability and curiosity. Sociability uses a linear approach to group shared competences to approach problems, whereas curiosity uses a non-linear approach. When the parameter curiosity is in place (i.e. set to "on"), senior developers are more effective in combining team members' competence into the task to be performed or problem to be solved. This implies that the performance on a given story depends on the senior developer that finds a connection to that particular story and the team he/she forms with other developers for the task. The interpretation of results is that senior developers with only sociability but no curiosity work better in a team when the team members have something to offer (are also competent), and less inclined to share when the "quality" (competence) of the team is low. For this individual tendency to be overcome, senior developers need to have both sociability and curiosity.

## 5. IMPLICATIONS AND CONCLUSIONS

The implication to practice is that what counts in an organization is not necessarily the single piece of expertise that each individual may bring in. Conversely, what seems to count is more related to the attitude that each individual has in relation to the problems the organization is facing. This is particularly applicable to senior developers in the team. So, care should be taken in hiring senior developers with behavior interviews that assess both sociability and curiosity traits. It is the open-mindedness and curiosity of the developers that may help cross competence gaps and subsequently bring about new and innovative perspectives, which would be simply unthinkable in the rigid confinements of static competence levels. However, an empirical study of a real Agile team is required to cement the understanding of how this may actually happen in practice.

The curiosity trait may be relevant to organizations that face a crisis or operate in turbulent environments. Employees that are capable of crossing the norms of their group and of integrating knowledge on a wider basis can make a better use of human resources. Here again, this may suggest that curious people adapt better to uncertainty, and they are consequently a better fit in turbulent environments because that is not after all so far from their usual way of working. Agile Scrum teams with sociable and curious members seem to find ways to increase their competence more rapidly than other teams in the organization. This is because of the specific assumption we have used to generate the model that sociable and curious team members collaborate better with each other, regardless of seniority level and individual goals, to find new solutions to complex problems. The best solution does not always come from the senior developers in the team working on their own. It most often comes when senior developers work collaboratively with the junior developers in their team as well as reach out to other teams in the Agile organization. As more senior developers evidence this fact, they become more inclined to be sociable and curious.
There are some limitations in the approach taken to study the proposed model of rational decision-making. First, competence c is represented by a number and that is an extremely simplified version of the way knowledge actually materializes. This calls for expanded and more sophisticated modeling. Second, the model only considers a constant number of developers and stories. The effect of increasing the number of stories and/or developers on team velocity can be the focus of a future study. Third, the effect of difficulty and mutation levels of stories on the time to develop a similar level of team competence needs to be explored. Fourth, additional factors like organization structure, rewards, stakeholder influence and leadership style should be incorporated in an improved version of the model. Finally, there are several parameters in this rich model that this study has not made use of and the hope is that they will be put to use in future enquiries by other researchers.

ACKNOWLEDGEMENTS

The author would like to thank Emanuele Bardone and Davide Secchi for providing the guidance and source ABM model for this study. This study uses a modification of the Agent-based model developed by them and is strongly motivated by the paper "Bardone, E. and Secchi, D. (2017), "Inquisitiveness: distributing rational thinking", Team Performance Management, Vol. 23 No. 1/2, pp. 66-81. https://doi.org/10.1108/TPM-10-2015-0044.

REFERENCES

[1] Bardone, E. and Secchi, D. (2017). "Inquisitiveness: distributing rational thinking", Team Performance Management, Vol. 23 No. 1/2, pp. 66-81. https://doi.org/10.1108/TPM-10-2015-0044.
[2] Bardone, E. (2011), Seeking Chances: From Biased Rationality to Distributed Cognition, volume 13 of Cognitive Systems Monographs, Springer, New York, NY.
[3] Cannon-Bowers, J.A. and Salas, E. (2001). "Reflections on shared cognition", Journal of Organizational Behavior, Vol. 22 No. 1, pp. 195-202.
[4] Cannon-Bowers, J.A., Salas, E. and Converse, S. (1993), "Shared mental models in expert team decision making", in Castellan, N.J. (Ed.), Individual and Group Decision Making, Lawrence Erlbaum, Hillsdale, NJ, pp. 221-246.
[5] Chia, R. and Holt, R. (2009), Strategy Without Design: The Silent Efficacy of Indirect Action, Cambridge University Press, Cambridge.
[6] Cowley, S.J. and Vallee-Tourangeau, F. (Eds) (2013), Cognition beyond the Brain: Computation, Interactivity and Human Artifice, Springer, London.
[7] Croissant, Y. and Millo, G. (2008), "Panel data econometrics in r: the plm package", Journal of Statistical Software, Vol. 27 No. 2.
[8] Ferdinand, A.E. (1974), A Theory of System Complexity, International Journal of General Systems, 1:1, 19-33, DOI: 10.1080/03081077408960745.
[9] Fioretti, G. (2013), "Agent-based simulation models in organization science", Organizational Research Methods, Vol. 16 No. 2, pp. 227-242.
[10] Gigerenzer, G. and Selten, R. (2001), Bounded Rationality: The Adaptive Toolbox, MIT Press, Cambridge, MA.
[11] Gilboa, I. (2010), Rational Choice, first edition, MIT Press, Cambridge, MA.
[12] Greene, W.H. (2008), Econometric Analysis, 6th ed., Prentice Hall, Upper Saddle River, NJ.
[13] Hutchins, E. (1995), Cognition in the Wild, MIT Press, Cambridge, MA.
[14] Kahneman, D. (2003), "A perspective of judgement and choice: mapping bounded rationality", American Psychologist, Vol. 58 No. 9, pp. 697-721.
[15] Klein, K. and Kozlowski, S. (2000), "From micro to meso: critical steps in conceptualizing and conducting multilevel research", Organizational Research Methods, Vol. 21 Nos 1/2, pp. 211-236.
[16] Knudsen, T. (2003), "Simon's selection theory: why docility evolves to breed successful altruism", Journal of Economic Psychology, Vol. 24 No. 2, pp. 229-244.
[17] Knudsen, T. and Levinthal, D.A. (2007), "Two faces of search: alternative generation and alternative evaluation", Organization Science, Vol. 18 No. 1, pp. 39-54.
[18] Kuhn, T. (2009), The Structure of Scientific Revolutions, 4th ed., Chicago University Press, Chicago.
[19] Lalsing, V. (2012). People Factors in Agile Software Development and Project Management. International Journal of Software Engineering & Applications, 3(1), 117–137. https://doi.org/10.5121/IJSEA.2012.3109.
[20] Lorscheid, I., Heine, B.O. and Meyer, M. (2012), "Opening the 'black box' of simulations: increased transparency and effective communication through the systematic design of experiments", Computational and Mathematical Organization Theory, Vol. 18 No. 1, pp. 22-62.
[21] Magnani, L. (2007), Morality in A Technological World: Knowledge as a Duty, Cambridge University Press, New York.
[22] Magnani, L., Secchi, D. and Bardone, E. (2007), "The docile hacker: the open source model as a way of creating knowledge", in Schmidt, C. (Ed.), Proceedings from the International Conference on Computers and Philosophy, i-C&P, Laval, pp. 583-596.
[23] March, J.G. (1994), A Primer on Decision Making, Free Press, New York, NY.

International Journal of Software Engineering & Applications (IJSEA), Vol.13, No.5, September 2022[24] McAvoy, J., & Butler, T. (2007). The impact of the Abilene Paradox on double-loop learning in an agile team. Information and software technology, 49(6), 552-563.

[25] Meslec, N. and Graff, D. (2015), "Being open matters: the antecedents and consequences of cross-understanding in teams", Team Performance Management, Vol. 21 Nos 1/2, pp. 6-18.

[26] Miller, K.D. and Lin, S.J. (2010), "Different truths in different worlds", Organization Science, Vol. 21 No. 1, pp. 97-114.

[27] Ossola, P. (2013), "Trust as a mechanism to increase 'docility.' a theoretical approach", International Journal of Organization Theory and Behavior, Vol. 16 No. 4, pp. 495-520.

[28] Polhill, J.G. (2010), "ODD updated", Journal of Artificial Societies and Social Simulation, Vol. 13 No. 4, p. 9.

[29] Radax, W. and Rengs, B. (2010), "Prospects and pitfalls of statistical testing: insights from replicating the demographic prisoner's dilemma", Journal of Artificial Societies and Social Simulation, Vol. 13 No. 4, p. 1.

[30] Rivkin, J.W. (2000), "Imitation of complex strategies", Management Science, Vol. 46 No. 6, pp. 824-844.

[31] Secchi, D. (2011), Extendable Rationality: Understanding Decision Making in Organizations, Springer, New York, NY.

[32] Secchi, D. (2015), "A case for agent-based model in organizational behavior and team research", Team Performance Management, Vol. 21 Nos 1/2, pp. 37-50.

[33] Secchi, D. (2016), "Boundary conditions for the emergence of 'docility:' An agent-based model and simulation", in Secchi, D. and Neumann, M. (Eds), Agent-Based Simulation of Organizational Behavior: New Frontiers of Social Science Research, Springer, New York, NY, pp. 175-200.

[34] Secchi, D. and Adamsen, B. (2017), "Organizational cognition: a critical perspective on the theory in use", in Cowley, S.J. and Vallee-Tourangeau, F. (Eds), Cognition Beyond the Brain: Computation, Interactivity and Human Artifice, 2nd ed., Springer, Heidelberg.

[35] Secchi, D. and Bardone, E. (2009), "Super-docility in organizations: an evolutionary model", International Journal of Organization Theory and Behavior, Vol. 12 No. 3, pp. 339-379.

[36] Secchi, D. and Bardone, E. (2013), "Socially distributed cognition and intra-organizational bandwagons: theoretical framework, model, and simulation", International Journal of Organization Theory and Behavior, Vol. 16 No. 4, pp. 521-572.

[37] Secchi, D. and Gullekson, N. (2016), "Individual and organizational conditions for the emergence and evolution of bandwagons", Computational and Mathematical Organization Theory, Vol. 22 No. 1, pp. 88-133.

[38] Secchi, D. and Neumann, M. (Eds) ( 2016), Agent-Based Simulation of Organizational Behavior: New Frontiers of Social Science Research, Springer, New York, NY.

[39] Secchi, D. and Seri, R. (2016), "Controlling for 'false negatives' in agent-based models: a review of power analysis in organizational research", Computational and Mathematical Organization Theory, Vol. 23 No. 1.

[40] Seri, R. and Secchi, D. (2017), "The problem of determining the number of runs in your simulation", in Edmonds, B. and Meyer, R. (Eds), Simulating Social Complexity. A Handbook, 2nd ed., Springer, Heidelberg.

[41] Siggelkow, N. and Rivkin, J.W. (2005), "Speed and search: designing organizations for turbulence and complexity", Organization Science, Vol. 16 No. 2, pp. 101-122.

[42] Simon, H.A. (1955), "A behavioral theory of rational choice", Quarterly Journal of Economics, Vol. 69 No. 1, pp. 99-118.

[43] Simon, H.A. (1990), "A mechanism for social selection and successful altruism", Science, Vol. 250 No. 4988, pp. 1665-1668.

[44] Simon, H.A. (1993), "Altruism and economics", American Economic Review, Vol. 83 No. 2, pp. 156-161.

[45] Simon, H.A. (1997), Administrative Behavior, 4th ed., The Free Press, New York, NY.

[46] Thomsen, S.E. (2016), "How docility impacts team efficiency: an agent-based modeling approach", in Secchi, D. and Neumann, M. (Eds), Agent-Based Simulation of Organizational Behavior: New Frontiers of Social Science Research, Springer, New York, NY, pp. 159-173.

[47] von Neumann, J. and Morgenstern, O. (1944), Theory of Games and Economic Behavior, Princeton University Press, Princeton, NJ.

[48] Wilensky, U. (1999), Netlogo: Center for Connected Learning and Computer-Based Modeling, Northwestern University, Evanston, IL.
11

## APPENDIX

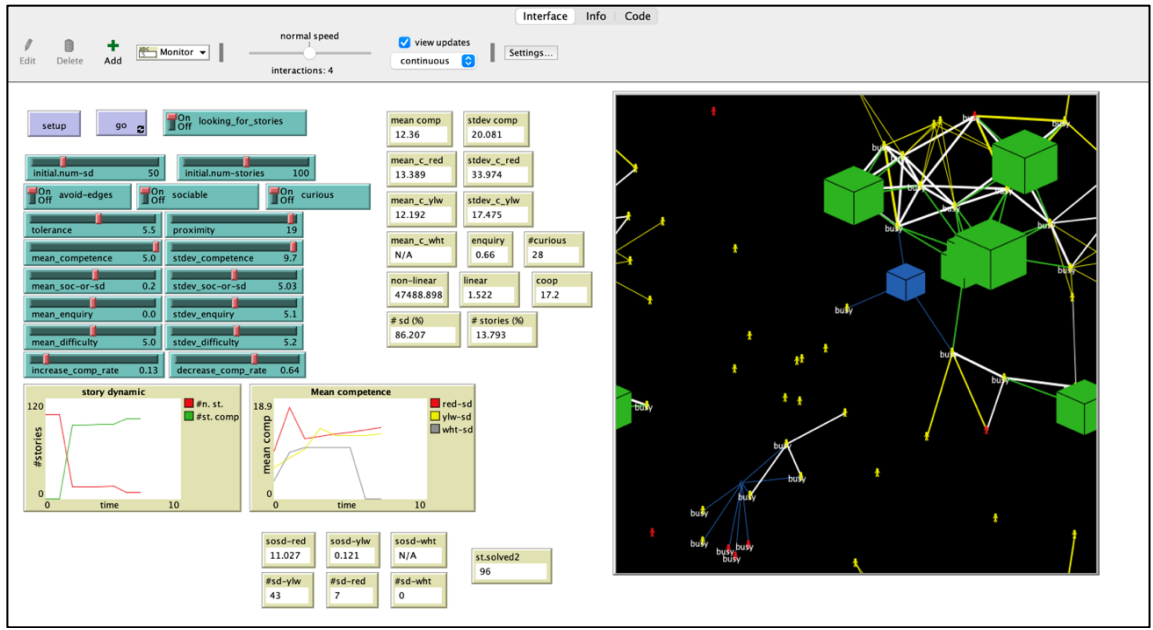

Figure 1. NetLogo 6.2.2 simulation environment.

## AUTHOR

**Ravi kalluri**, MSME, MSEE, MBA, PMP, ACP, SPT, CSM, Ravi has 25 years of experience leading complex programs, with an expertise in project risk management. Currently, he serves as a Sr. Technical Program Manager at Amazon Web Services. He has previously held program management roles at Cisco Systems, Citrix Systems, Ericsson, CA Technologies, VeriSign, and Motorola. Ravi previously served as an Adjunct Faculty of Project Management in the University of California System (Berkeley, Irvine, Santa Cruz Extension campuses). He has conducted corporate trainings in Project 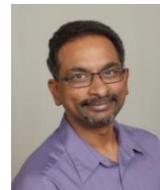
Management for Volkswagen of America as well as taught project management and risk management courses for Northrop Grumman. His recent presentations include Project Risk Management for Business Success in a Startup Environment at PMI Silicon Valley and Enabling Collaborative Software Development in a Post-Acquisition Scenario at VeriSign Technical Symposium. Ravi received his MBA from Northwestern University (Kellogg), MS in Electrical Engineering at Stanford, MS in Mechanical Engineering at The Ohio State University, MS Certificate in Project Management from George Washington University and Graduate Certificate in Systems Engineering from Massachusetts Institute of Technology.